\documentclass[letterpaper]{article}
\pdfoutput=1
\usepackage{jheppub}
\newcommand{\mhlf}{{m_{\frac{1}{2}}}}
\newcommand{\msq}{{M_{\widetilde q}}}
\newcommand{\mgl}{{M_{\widetilde g}}}
\newcommand{\mchizo}{M_{\widetilde\chi^0_1}}
\newcommand{\mno}{M_{\widetilde{N}_1}}
\newcommand{\none}{\widetilde{N}_1}
\newcommand{\cone}{\widetilde{C}_1}
\newcommand{\ntwo}{\widetilde{N}_2}
\newcommand{\nsubi}{\widetilde{N}_i}
\newcommand{\nj}{\widetilde{N}_j}
\newcommand{\ci}{\widetilde{C}_i}
\newcommand{\cj}{\widetilde{C}_j}
\newcommand{\sq}{{{\widetilde q}}}
\newcommand{\gl}{{{\widetilde g}}}
\newcommand{\met}{\slash\!\!\!\!{E}_T}
\newcommand{\mht}{\slash\!\!\!\!{H}_T}

\title{A Closer Look at the 2011 cMSSM Results from CMS}
\author{Stephen Mrenna}
\affiliation{SSE Group,\\Computing Division,\\ Fermilab,\\ Batavia, IL  60510}
\date{\today}
\emailAdd{mrenna@fnal.gov}

\abstract{
We present a phenomenological appraisal of the results of several
searches for Supersymmetry ({\sc SUSY}) performed
at the LHC by the CMS collaboration
and interpreted in the context of the cMSSM.
Part of the analysis focuses on which {\sc SUSY} production processes are 
being probed.  We observe that much of the current exclusion region
is dominated by squark-squark and squark-gluino production,
and explain the shape of the exclusion curves.
Based on this analysis and an estimation of future reach, additional simplified models are 
suggested.
Other phenomenological details are discussed, such as the effect of
radiation on acceptance.
}

\begin{document}
\mbox{FERMILAB-PUB-11-551-CD}
\maketitle

\section{Introduction}

\begin{figure}[!ht]
\label{fig:cmsSUSYCMSSM}
\begin{center}
\includegraphics[width=0.85\textwidth]{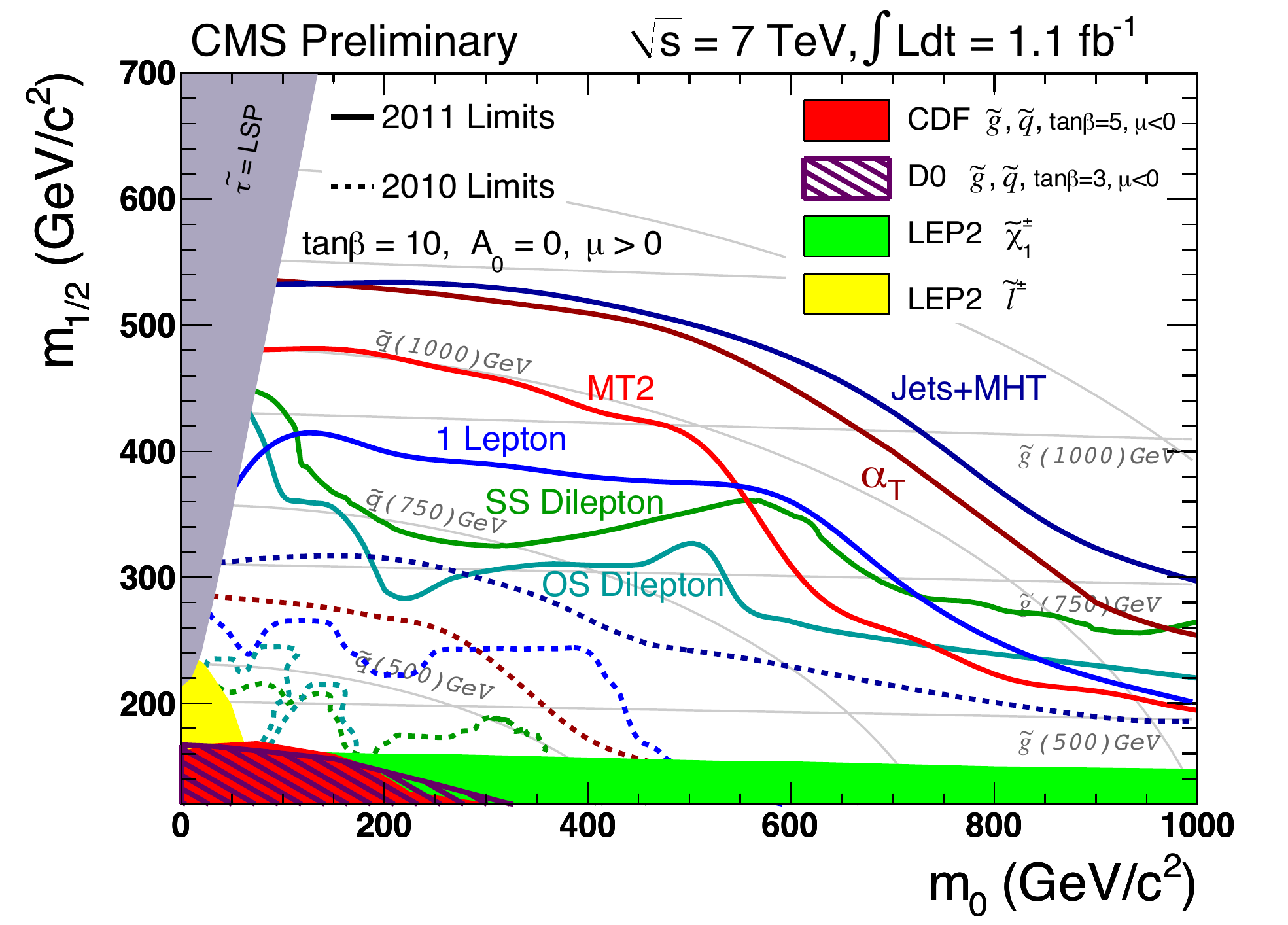}
\caption{
Observed limits from several 2011 CMS SUSY searches plotted in the cMSSM 
($m_0,\mhlf$) plane, with $\tan\beta=10, A_0=0, sgn(\mu)>0$.
}
\end{center}
\end{figure}

Significant effort at the LHC is dedicated to 
analyzing data to find evidence of, or to exclude, models
of Supersymmetry {(\sc SUSY)}
(Refs.~\cite{Martin:1997ns,Chung:2003fi} provide thorough theoretical reviews.) 
Many of the results of these analyses are interpreted
in the context of the constrained Minimal Supersymmetric
Standard Model {\sc (cMSSM)}~\cite{Kane:1993td}, with exclusion contours
shown in the plane of the universal scalar mass $m_0$ and
the gaugino mass $\mhlf$ parameters, with other {\sc cMSSM} parameters fixed ($\tan\beta,A_0,sgn(\mu)$).
A recent result from CMS \cite{cmsSummary} is presented in Fig.~\ref{fig:cmsSUSYCMSSM},
based on 1.1 fb$^{-1}$ of data.

Our initial attention is drawn to the curves in Fig.~\ref{fig:cmsSUSYCMSSM}
labeled
{\it{Jets+mHT}}~\cite{CMS:RA2}, 
{\it{$\alpha_T$}}~\cite{Collaboration:2011zy}, and 
{\it{MT2}}~\cite{CMS:MT2}, referring to limits from specific
analyses known by those colloquial names.  These analyses
are sensitive mainly to the production and 
hadronic decays of sparticles, whereas the other analyses,
with limit contours labeled {\it 1 Lepton}, {\it SS/OS Dilepton}\cite{CMS:OneL,CMS:SS,CMS:OS}, 
depend explicitly on
branching fractions of sparticles to leptons (either directly or through decays of gauge bosons).

The hadronic limits demonstrate that:
\begin{itemize}
\item for low $m_0$, there is a significant reach in $\mhlf$ ($\sim 540$ GeV);
\item for large $m_0$, $\mhlf$ is limited to smaller values ($\sim 300$ GeV).
\end{itemize}
The parameters $m_0$ and $\mhlf$ are {\it not} 
the physical masses of the theory, which set the size of production cross sections
and the kinematics of decays.
However, $\mgl$ is basically set by $\mhlf$ ($\mgl \sim 2.5\mhlf$), while
$\msq^2$ has a quadratic dependence on $m_0$ and $\mhlf$.
From Fig.~\ref{fig:cmsSUSYCMSSM}, it is clear
that the excluded gluino mass  is
not a flat line: the exclusion depends strongly on the squark mass.
(Figure~\ref{fig:squark-gluino} of the Appendix shows the correlations between 
$\mgl$ and $\msq$ in the ($m_0,\mhlf$) plane relevant for this analysis.)
In particular, one can see from the contours of fixed gluino
and squark mass, labeled $\sq$ and $\gl$, that $\msq<$ 1 TeV
is excluded regardless of $\mgl$, whereas $\mgl<$ 1 TeV is only excluded for
$\msq<$1.1 TeV.  The exclusion curve seems to follow the semi-circular contours
of fixed $\msq$ for low $m_0<1~{\rm TeV}$, and the flat contours of fixed $\mgl$ for
larger $m_0$.

The physical masses of the cMSSM are highly correlated, as expected
from a model that has many observables but only a few parameters.
While the quantitative structures of the cMSSM depend on the
exact specification of $m_0,\mhlf,A_0,\tan\beta,sgn(\mu)$, the
qualitative features are set by $m_0$ and $\mhlf$ (for larger values
of $\tan\beta$, third generation sparticles may be lighter.  
From the collider physics phenomenology point of view, this
affects the other spectra mildly, but can have significant implications
for decay modes.)
It is well known that $\msq > \mgl$ in most of this plane and that
the gaugino masses come in specific ratios
($\mno \equiv \mchizo \sim \frac{1}{6}\mgl$).
One can question whether general results can be learned from examining
the limits within the {\sc cMSSM}, and for a single choice of {\sc cMSSM}
parameters.
In particular, this limit is for $\tan\beta=10, A_0=0~{\rm GeV}, sgn(\mu)>0$.
(We also comment on the case of $\tan\beta=40, A_0=-500~{\rm GeV}, sgn(\mu)>0$.)
We argue that several lessons can be learned from this particular
cMSSM interpretation, despite the fact that it is explicitly model-dependent.

In this paper, we will investigate:
\begin{enumerate}
\item the dominant SUSY processes that contribute to these limits.  In particular, we would
like to understand the shape of the CMS exclusion curves in Fig.~\ref{fig:cmsSUSYCMSSM}.  Analyzing
  the plane this way is a mapping of a point in $(m_0,\mhlf)$ into simple or simplified models;
\item the impact of this mapping on the selection of simplified models for future studies to better
 understand the experimental results;
\item the sensitivity of the various analyses to specific {\sc SUSY} production processes and decay modes;
\item phenomenological details relevant to this
analysis and other new particle searches, such as kinematic features of
various processes and the role of radiation in searches.
\end{enumerate}

\section{Dominant Processes}

\begin{figure}[!ht]
\begin{center}
\includegraphics[width=0.7\textwidth]{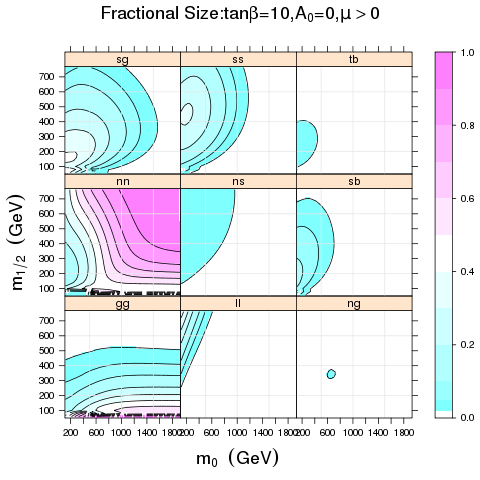}
\caption{
Fractional size of various subprocesses as a function
of ($m_0,\mhlf$) for $\tan\beta=10, A_0=0~{\rm GeV}, sgn(\mu)>0$.
}
\label{fig:frac2}
\end{center}
\end{figure}

The purpose of this section is to examine which processes
are contributing to our current exclusion contours in the cMSSM.
Here, we concentrate on the size of production cross sections without
consideration of a specific selection of events, as in a search analysis.
We will justify this later.
The deconstruction of the cMSSM into subprocesses is a mapping into
a set of simplified model spectra (SMS) (see \cite{Alves:2011wf}, for example).
This will help us to understand the generality of these cMSSM results,
and interpret them in simplified models.  The main result is presented
in Fig.~\ref{fig:frac2}, which shows the fractional importance of various
{\sc SUSY} pair production processes as a function of
$(m_0,\mhlf)$ for the same parameters as in Fig.~\ref{fig:cmsSUSYCMSSM}.

The labels in Fig.~\ref{fig:frac2} represent (from top-left to bottom-right)
the following production processes:
{\it sg} = squark-gluino, 
{\it ss} = squark-squark, 
{\it tb} = stop-antistop,
{\it nn} = neutralino/chargino pair,
{\it ns} = neutralino/chargino-squark,
{\it sb} = squark-antisquark,
{\it gg} = gluino pair,
{\it ll} = slepton-antislepton,
{\it ng} = neutralino/chargino-gluino.
The classifications sum over similar processes: {\it nn}, for example, 
includes all $\nsubi\nj,\ci\cj,\nsubi\cj$ processes.   (In practice, only a
few processes tend to saturate the sum, e.g. $\cone\ntwo$ and $\cone\cone$ 
in this case.)
For brevity, the mostly-negligible sbottom-antisbottom pair production is
not shown in Fig.~\ref{fig:frac2}.  These and other production cross sections were calculated
using the computer code {\tt Prospino} to NLO accuracy \cite{Prospino}.
The physical particle properties were calculated using the
MSSM evolution code 
{\tt SoftSUSY}~\cite{Allanach:2001kg}, while sparticle and Higgs boson
decays widths were calculated from
{\tt SUSYHIT}~\cite{Djouadi:2006bz} (sparticle decays are considered later).

The all-hadronic searches 
naively target squark and gluino production.
Figure~\ref{fig:frac2} demonstrates that
squark-squark production
and squark-gluino production have large rates at small $m_0$.
For small $\mhlf$, gluino pair production is important, while
the production of electroweak gauginos is important,
if not dominant,
over most of the plane.

\begin{figure}[!ht]
\begin{center}
\includegraphics[width=0.7\textwidth]{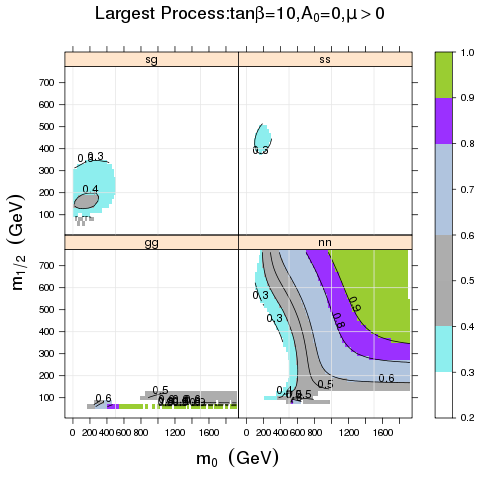}
\caption{
The largest cross section in the ($m_0,\mhlf$) plane,
with contours showing the fractional size.
}
\label{fig:frac3}
\end{center}
\end{figure}

For further clarity, we focus on the largest process
at each point in the plane, which is presented 
in Fig.~\ref{fig:frac3}. 
Only four classes of processes every satisfy the
criterion of being the largest cross section
at any point (a small region of
slepton production for small $m_0$ and large $\mhlf$ is not shown).
When squark-squark production is the largest
(roughly $400<m_0<500~{\rm GeV}, 100<\mhlf<300~{\rm GeV}$), 
it comprises 30-40\% of the total {\sc Susy} rate
at that point.\footnote{The large contribution of {\it ss} processes
to the total {\sc SUSY} production cross section was noted for
the LM1 benchmark point in Ref.~\cite{Khachatryan:2011tk}.}
Gluino-pair production is largest in
only a limited region of $\mhlf$ (roughly
$\mhlf<120~{\rm GeV}$); the other gauginos have much more significant production rates.
Clearly, searches that are sensitive to the
electroweak gaugino processes 
(mainly $\widetilde{C}^\pm_1\widetilde{C}^\mp_1, \widetilde{C}^\pm_1\widetilde{N}_2$) can
gain inroads to the upper-right portion
of the plane.  {\it sg} is competitive with {\it ss}, and the largest process for
a the region $m_0<400{~\rm GeV}, 100<\mhlf<300{~\rm GeV}$.
We reproduce Fig.~\ref{fig:frac2} for the case of large $\tan\beta$,
namely $\tan\beta=40, A_0=-500~{\rm GeV}, sgn(\mu)>0$, in Fig.~\ref{fig:tanb40}
of the Appendix.  The qualitative behavior is the same.

These conclusions are noteworthy.  
First, squark-squark production can only occur if a
gluino exists, whereas squark-antisquark production
arises directly from the QCD Lagrangian.   
Even if the gluino is heavy enough to have a suppressed
production cross section, the
squark-squark production cross section can be 
substantial (we study this more quantitatively later).
Thus, an indirect way to prove or disprove the existence
of a gluino with a TeV-scale mass is to establish that a signal
is squark-squark or squark-antisquark production.
Potential clues could be obtained through jet-charge
or exploiting some differences in gluon radiation
arising from the contributions of valence or sea
quarks to the process (see later discussion).

Squark-gluino production, squark-squark, and
gluino-gluino production processes are all competing at once.
We will investigate what mass relations between
the squark and gluino allow squark-gluino production
to dominate over the other possible production processes.

The {\it nn} processes are mainly
the {\sc SUSY} duals of $WW$ and $WZ$ production, which is
typical of the cMSSM.
Almost all of {\it nn} is $\widetilde{C}_1^\pm \widetilde{N}_2$ and 
$\widetilde{C}_1^\pm \widetilde{C}_1^\mp$ production, 
with the lightest chargino ${\widetilde C}_1$ and the second heaviest neutralino 
${\widetilde N}_2$ nearly degenerate 
in mass.

The cross sections for stop and sbottom pair production are never
competitive with other processes (even with
$\tan\beta=40$) for these {\sc cMSSM} parameters.  We will discuss this more later.

\begin{figure}[!ht]
\begin{center}
\includegraphics[width=0.75\textwidth]{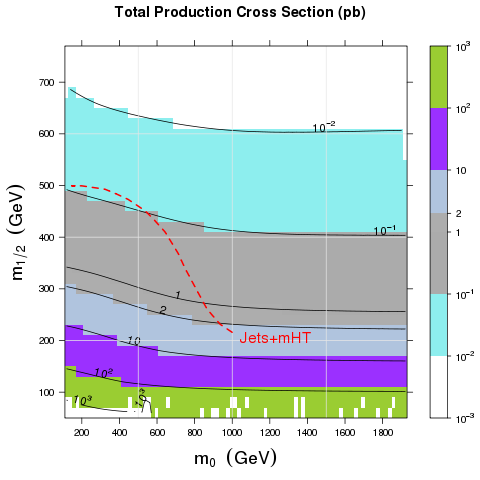}
\caption{
The total {\sc SUSY} production cross section as a function
of ($m_0,\mhlf$) in the cMSSM for $\tan\beta=10, A_0=0~{\rm GeV}, sgn(\mu)>0$.
The strongest limit curve from Fig.~\ref{fig:cmsSUSYCMSSM} is overlaid with
the label {\it Jets+mHT}.
}
\label{fig:sigmatotal}
\end{center}
\end{figure}

The
{\it total} {\sc SUSY} production cross section 
in the $(m_0,\mhlf)$ plane 
is presented in Fig.~\ref{fig:sigmatotal}.
The contours of constant cross section are nearly horizontal
lines in $\mhlf$.  This is not surprising, since the cross section
for the {\it nn} process is set mainly by the chargino and neutralino masses.
Comparing with the limits in Fig.~\ref{fig:cmsSUSYCMSSM} (and reproduced
in this figure), we observe that
current searches with 1.1 fb$^{-1}$ of data are excluding 
the cMSSM points that yield a total cross section 
of several hundred fb, or several hundred {\sc Susy} events
before event selection.

\section{Simplified Models}

The simplified model approach originated as a topological
classification of signals to allow hypothesis testing
\cite{Knuteson:2006ha,ArkaniHamed:2007fw}.  It is based on the observation
that the kinematics of new particle production and decay
is often characterized by the mass scales involved, and
not by the particulars of matrix elements.   In the
absence of signals, simplified models (SMS) can be used
to present results with less model dependence~\cite{Alves:2011wf}.
Thus, {\sc SUSY} search limits can be recycled to
provide information to the community about the dependence of
these limits on the masses and decay chains in a particular
model.  SMS are also useful to the analysts, since
it reveals the coverage of
the cuts used in a particular analysis.

By breaking down the results in the cMSSM into individual processes,
we are mapping into simplified models.  Do we have the right ones or
enough? 

This simple analysis of only the production cross sections
can already have an impact on the choice of
simplified models used to interpret the data.
The main suggested topologies, $\gl\gl$ and 
$\sq\sq^*$ \cite{Alwall:2008ag}, are not the
most relevant in most of the cMSSM plane, even though we have relied
on these processes to motivate our initial choices of
topologies.   
If our aim is to provide motivated SMS that
cover the main possibilities in a model like the cMSSM, and/or
to allow for the deconstruction of a particular point in 
the cMSSM plane in terms of SMS, then
other topologies need to be considered, such as:
\begin{enumerate}
\item squark-squark ({\it ss}) production;

\item squark-gluino ({\it sg}) production;

\item Weak gaugino ({\it nn}) production, 
like $\cone^\pm\cone^\mp$ and $\cone^\pm\ntwo$.   
\end{enumerate}

However, there are caveats.
A specific point
in the $(m_0,\mhlf)$ plane will likely not have the simple structure
of the simplest of SMS, with a single decay chain.
Instead, cascade decays like
$\sq_R \to q \ntwo, \sq_L \to q \ntwo, q \cone$ are competing with
single step decays like $\sq\to q\none$.
In general, cascade decays tend to soften the $\met$ spectrum, but can provide
more jets that increase acceptance.
Thus, limits in the SMS may be different than those in
the cMSSM for similar choices of $\msq,\mgl,\mno$.

First, we examine the $\sq\sq$ process 
and compare the kinematics of $\sq\sq$
and $\sq\sq^*$ production.
We expect some differences from the effect of the parton distribution
functions, since valence-valence quark scatterings can only contribute to $\sq\sq$ production.
Differences arise at
the hard-process level, but also from initial state radiation.

\begin{figure}[!h]
\begin{center}
\includegraphics[width=0.75\textwidth]{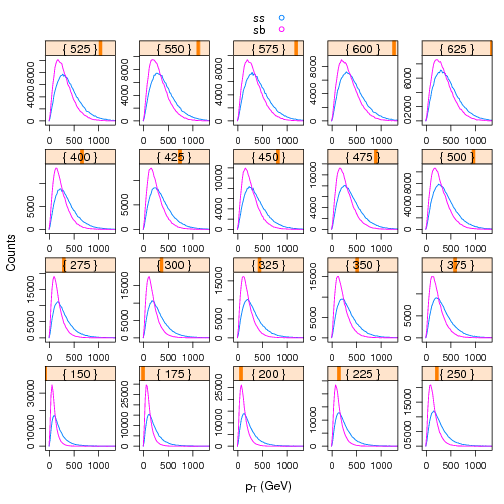} 
\caption{Comparison of the transverse momentum $p_T$ 
of the produced $\sq,\sq^*$.  Plots of rapidity $y$ are shown
in the Appendix.}
\end{center}
\label{fig:sssbkine}
\end{figure}

Figure~\ref{fig:sssbkine} shows comparisons of the $\sq$ (or $\sq^{*}$) 
$p_T$ after the hard event and parton shower generation from
{\tt Pythia} (see the Appendix for details).   
Comparisons of the rapidity are also shown in the Appendix.
Values of $150\le\msq\le 650~{\rm GeV}$ are displayed, though
we have studied larger values.
In general, differences are less pronounced for larger $\msq$.
For $\msq < 800~{\rm GeV}$, 
the $p_T$ of the squark is harder for {\it ss} production, while the
rapidity $y$ is less central.  Thus, the acceptance of {\it ss} events should
be different from {\it sb}.  If the acceptance is also dependent upon a
jet arising from initial state radiation, there will be further differences.

Since interpretations of data with simplified models have been presented only for 
$\gl\gl$ and $\sq\sq^*$ topologies
so far (see, for example, Ref.~\cite{Collaboration:2011ida}), 
we encourage a study of $\sq\sq$ topologies.
However, as a first estimate, one can ignore the differences in acceptance,
and interpret the data using {\it ss} cross sections, instead of {\it sb}.
This requires a choice of $\mgl$, since the {\it ss} cross section
vanishes otherwise.   We have checked that the kinematic dependence
on $\mgl$ is negligible, and the $\sq\sq$ acceptance for $\mgl=2~{\rm TeV}$
is nearly identical to that for $\mgl=4~{\rm TeV}$, so that 
signal events can be studied ``independently'' of $\mgl$.   
Later, we will see
that the {\it ss} cross section exceeds the {\it sb} one even for 
cases when $\mgl$ is quite large.

\begin{figure}[!ht]
\begin{center} 
\includegraphics[width=0.75\textwidth]{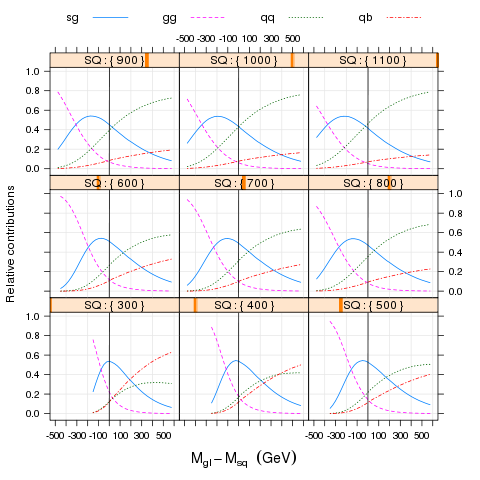}
\caption{
The relative {\sc SUSY} QCD production cross sections as a function
of the mass splitting between the gluino and squark.
The vertical line is for degenerate $\msq,\mgl$.
}
\label{fig:squarkgluino}
\end{center}
\end{figure}

Next, we turn to the squark-gluino process.
We consider the complete MSSM description of 
squarks and gluinos, so that all production cross sections
are competing and interconnected.  We have discussed already
the $\gl\gl$, $\sq\sq$, and $\sq\sq^*$ contributions.
In Figure~\ref{fig:squarkgluino}, we show the relative 
size of each cross section as a function of
the mass splitting between the gluino and squark, for
several fixed values of $\msq$.
{\it sg} is the dominant production process in a small window,
roughly near $\mgl \sim \msq$,
that changes moderately with $\msq$.   
For low $\msq$, the dominance occurs for nearly degenerate
squarks and gluinos.   However, for higher $\msq$, the
maximum occurs when the gluino is somewhat lighter than
the squarks.  This leads to a parton-level signature of
$\widetilde{q}(\to q\gl(\to q\bar q \none)\widetilde{g}(\to q\bar q \none)$.
Of course, specific cuts might select {\it sg} over other topologies,
so it could be important even what it is not the largest cross section.
This suggests SMS with $\mgl$ within $\pm 200$ GeV of $\msq$.
We note also the rise of $\sq\sq$ over $\sq\sq^*$ production
once $\msq>500-600$ GeV.  For $\msq=600$ GeV and
$\mgl=1200$ GeV, the $\sq\sq$ production cross section
is roughly 0.5 pb.

\begin{figure}[!ht]
\begin{center} 
\includegraphics[width=0.75\textwidth]{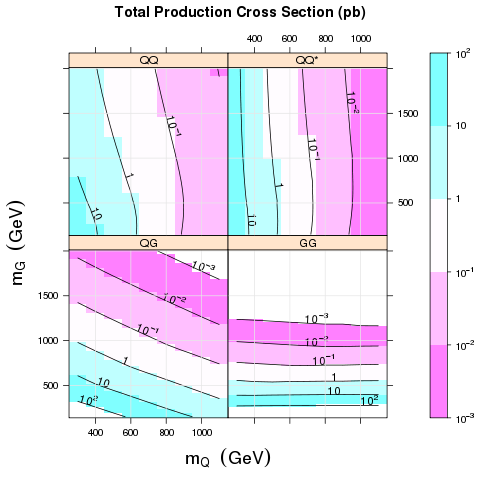}
\caption{
Levels of {\sc SUSY} QCD production cross sections as a function
of the gluino and squark masses.
}
\label{fig:squarkgluinolevels}
\end{center}
\end{figure}

In Figure~\ref{fig:squarkgluinolevels},
we present the same information as a contour plot in 
the $\msq,\mgl$ plane.
Note the unequal limits for each axes, chosen to 
emphasize the size of the $\sq\sq$ cross section for large
$\mgl$.  Focusing on the behavior of the $\sq\sq$ cross section
at fixed $\msq=400{\rm~GeV}$ as a function of $\mgl$,
one observes a decrease of only one order of magnitude in
the cross section as the $\mgl$ varies from 275 to 2000 GeV.

Next, we comment on electroweak gaugino pair production.
In the cMSSM, the mass splitting between the chargino
($\sim 0.8\mhlf$) and the LSP ($\sim 0.4\mhlf$) is roughly
$0.4\mhlf$.  Once $\mhlf>M_W/0.4\sim 200$ GeV, the
decay $\cone\to W\none$ is allowed for on-shell particles.
From Fig.~\ref{fig:cmsSUSYCMSSM}, we observe that the current leptonic analyses are {\it already} limiting models of
this type, since their contours extend up to and beyond $\mhlf\sim 200{\rm~GeV}$.
(We have not studied in detail the shapes of the leptonic limits, but leave this
for a future study.)
This may be viewed as a worst-case in terms of
leptonic branching ratio (at specific points, $\widetilde{N}_2 \to \widetilde{N}_1 h(\to b\bar b)$ can occur.   We comment on this later).  
Even hadronic searches may be sensitive to
these processes, which naively yield 4 hard jets
and $\met$.  For heavy enough $\cone$ and $\ntwo$, the gauge bosons produced in decays may be significantly
boosted. 

\begin{figure}[!ht]
\begin{center}
\includegraphics[width=0.65\textwidth]{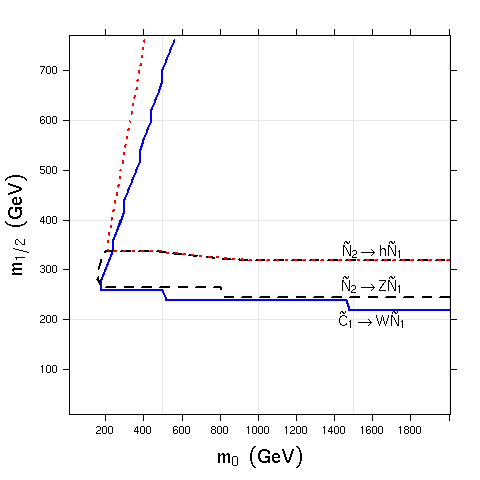}
\caption{
Regions where BR($C_1 \to W N_1$),
BR($N_2 \to Z N_1$), and
BR($N_2 \to h N_1$) are significant in the ($m_0,\mhlf$) plane
for $\tan\beta=10, A_0=0~{\rm GeV}, sgn(\mu)>0$.
}
\label{fig:ctown}
\end{center}
\end{figure}

In Figure~\ref{fig:ctown}, we show contours of significant decays
modes of $\cone$ and $\ntwo$ for $\tan\beta=10, A_0=0, sgn(\mu)>0$.  The contour of $BR(\cone\to W\none)\sim 1$
is the solid, black line near $\mhlf\sim 200{\rm GeV}$ -- our naive expectation
of where this decay should turn on.  The region where $\ntwo\to Z\none$
is significant is small and enclosed by the blue, long--dashed line, and its lower
bound is near the start of the $\cone$ contour.  Near $\mhlf\sim 300{\rm GeV}$,
a new decay mode is dominant, namely $\ntwo\to h\none$.  This has a (negative)
impact
on multi-lepton searches, but also opens the possibility of a $Wh\met$ signal.
The dominance of $\ntwo\to h\none$ is a feature, or peculiarity, of the
correlations inherit to a model like the cMSSM, with only a few input parameters.
A reasonable extension of the cMSSM is to decouple the Higgs scalar from the
sfermion scalar parameter.   A simple treatment, suitable for collider physics
phenomenology,  would be to remove the Higgs boson from the spectrum entirely,
and recalculate the decay widths of all sparticles.  A consistent approach to
this is handled in the {\sc pMSSM}~\cite{Sekmen:2011cz}.

Based on this, we suggest that a compelling simplified model is 
$\cone^\pm\cone^\mp$ and $\cone^\pm\ntwo$ production, with $\cone\to W^{(*)}\none$
and $\ntwo\to Z^{(*)}\none$ or $\ntwo\to h\none$.  The three-body decays to
off-shell $W$ or $Z$ bosons come into play when the mass splitting between
mother and daughter is small.   $M_{\cone} = M_{\ntwo}$ is a motivated choice, 
but $M_{\none}$ should be considered arbitrary. 
In the {\sc cMSSM}, there is model dependence arising from the $\sq$ contributions
in the matrix-element calculation of the production cross section.  The $\sq$ Feynman
diagrams
are much less significant (numerically) than the Standard Model gauge boson ones,
which arise naturally from the neutralino/chargino kinetic energy piece in the Lagrangian.
Reference cross sections can be chosen accordingly, ignoring $\sq$ contributions.
Because of the importance of gauge boson diagrams, $\cone^\pm\ntwo$ production
is roughly twice as large as $\cone^\pm\cone^\pm$ production, and $\cone^+\ntwo$ production
is roughly 60\% larger than $\cone^-\ntwo$ production.
This mirrors the behavior of $W^*$ and $\gamma^*/Z^*$ in the Standard Model.

\begin{figure}[!ht]
\begin{center}
\includegraphics[width=0.75\textwidth]{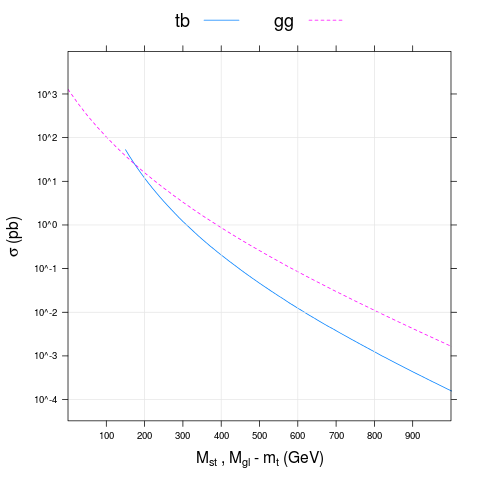}
\caption{ 
A comparison of stop-antistop and gluino pair production as a function
of mother mass.  The gluino mass $\mgl$ is offset by $m_t=175$ GeV, the
kinematic limit for the decay $\gl\to\widetilde{t}\bar{t}$.
}
\label{fig:stopVsgluino}
\end{center}
\end{figure}

Even though we have downplayed the pair production of heavy-flavor squarks,
this view may be too model dependent.   Contours of stop and sbottom mass
for the {\sc cMSSM} points considered here are shown 
in Fig.~\ref{fig:stop-sbottom-contour} of the Appendix.
A very large splitting between the stop and/or sbottom and the other squark
masses is not obtained (except for small $m_0$ and $\mhlf$) for these choices of {\sc cMSSM} parameters.
The stop is special in the MSSM, whether based on
fine-tuning arguments or its relation to the Higgs boson mass, and
one mass eigenstate may be significantly lighter than most other sparticles.
Later, we observe that the acceptance
for {\it tb} events is high for generic hadronic selections.
Targeted analyses, which may leverage 
the presence of $b$ or $t$ quarks in the final state, could have a much larger acceptance.
Here, we will consider
a scenario where $\none,\widetilde{t}$ and $\gl$ are the only relevant
sparticles in our current energy regime.
Figure~\ref{fig:stopVsgluino} shows the relation between gluino and
stop pair production as a function of mass, where the gluino mass has
been offset by the top mass (175 GeV) to emphasize the kinematic limit for $\gl\to\widetilde{t}\bar{t}$.
For a stop mass of 300(600) GeV, a gluino of mass 550(950) GeV
has a comparable production
cross section; for smaller mass splittings, gluino production will provide
more stop events, with additional $t$ quarks in the final state (half with the same sign).

\section{Studies of the {\it Jets+mHT} and $\alpha_T$ Analyses}

In this section, we justify our previous statement that an analysis
of the production cross section is meaningful.  To this end, we have
performed a generator-level analysis that replicates the
{\it Jets+mHT} and $\alpha_T$ event selections.   These two analyses are the
most straight-forward to reproduce (we leave $m_{T2}$, which
requires a hemisphere decomposition of events and an iteration
over invisible particle candidates, for a further study).
Details of the experimental analyses are to be found in the earlier references.
There are several results we wish to convey.  The first two demonstrate
that the current analyses are indeed sensitive to the dominant processes,
though in different ways.   A third shows the overlap between the two
analyses.  The last is to comment on the impact of a leptonic veto on these selections.
A discussion of the role of extra radiation of quarks
and gluons in the analysis results is presented in the next section.

\begin{figure}[!ht]
\begin{center}
\includegraphics[width=0.75\textwidth]{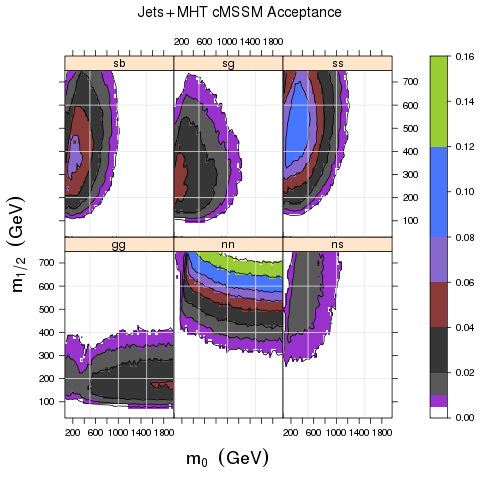}
\caption{
The acceptance (number of events that pass cuts divided by
the total number of events generated at the model point) broken down by
subprocess type for the {\it Jets+mHT} analysis.
}
\label{fig:classic1}
\end{center}
\end{figure}

Figure~\ref{fig:classic1} shows the acceptance of events in the $(m_0,\mhlf)$
plane, broken down by process type for the
{\it Jets+mHT} analysis.\footnote{We apply the Baseline cuts of $(H_T,\mht)>$(350,200) GeV.  
See the Appendix for other details.}  
Only six of the process classes are
significant.   For this plot, the acceptance is defined as the number
of events that pass the event selection per category, divided by the total
number of events studies for that value of $(m_0,\mhlf)$.   
The plot demonstrates that the {\it sg},
{\it ss}, and {\it nn} process classes are the most significant after applying the
selection cuts.  The contours of acceptance reflect the contours of
total cross section shown in Fig.~\ref{fig:frac2}.
This implies an acceptance that is relatively independent of process type,
provided that the primary sparticle masses are similar. 
The {\it nn} process exhibits a turn-on associated with the $\met$
threshold in the selection.

\begin{figure}[!ht]
\begin{center}
\includegraphics[width=0.75\textwidth]{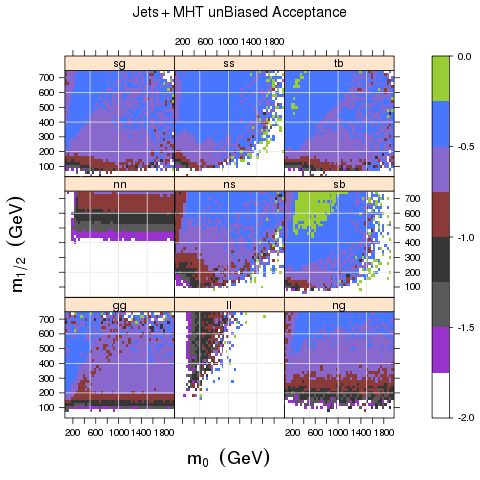}
\caption{
The acceptance (number of events that pass cuts divided by
the total number of events generated at the model point for
the given process) broken down by
subprocess type for the {\it Jets+mHT} analysis.
The contours are
drawn in a $\log_{10}$ scale.
}
\label{fig:classic2}
\end{center}
\end{figure}

A related quantity is the efficiency of an individual process
in isolation from the other processes in a full model, like the cMSSM.
This efficiency is the relevant one needed for a simplified model analysis.
Figure~\ref{fig:classic2} shows this unbiased acceptance in the
$(m_0,\mhlf)$ plane by processes type for the {\it Jets+mHT} selection.   
This analysis has a high efficiency for most types of processes, except
for the {\it nn} class, and {\it gg} for lower $M_{\gl}$.  
There are still correlations built
into these results; for example, the {\it ss} process depends upon the gluino mass.
However, the effect on kinematics is less than that on production rates,
as observed earlier.

The behavior of the {\it nn, gg, ng} acceptance is relatively easy to
interpret.   All the gaugino masses are roughly set by $\mhlf$ independent
of $m_0$, and the mass scales sets the hardness of the kinematics, and, hence,
the selection efficiency.  Too low values of $M_{\cone}, M_{\ntwo}$ will yield too
low $\met$ to pass these selections.
The {\it sg} acceptance is less obvious.
There is some loss of efficiency near the diagonal line.  From 
Fig.~\ref{fig:squark-gluino} in the Appendix, we observe that this
coincides with $\mgl\sim\msq$.  We expect that squark decays to
a soft jet and the gluino, and the gluino efficiency is lower because
of 3-body decays.

\begin{figure}[!ht]
\begin{center}
\includegraphics[width=0.75\textwidth]{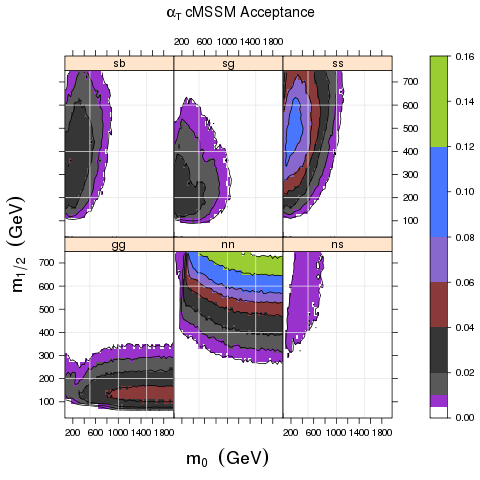}
\caption{
The acceptance (number of events that pass cuts divided by
the total number of events generated at the model point for
the given process) broken down by
subprocess type for the $\alpha_T$ analysis.
}
\label{fig:alfa1}
\end{center}
\end{figure}

\begin{figure}[!ht]
\begin{center}
\includegraphics[width=0.75\textwidth]{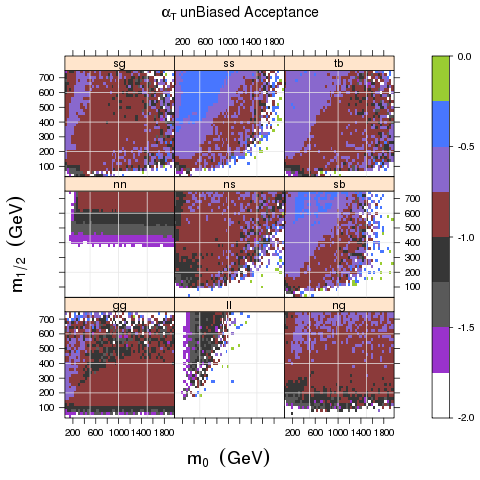}
\caption{
The acceptance (number of events that pass cuts divided by
the total number of events generated at the model point for
the given process) broken down by
subprocess type for the $\alpha_T$ analysis.  The contours are
drawn in a $\log_{10}$ scale.
}
\label{fig:alfa2}
\end{center}
\end{figure}

We have performed a similar analysis for the $\alpha_T$ event selection,
displayed in Figs.~\ref{fig:alfa1} and \ref{fig:alfa2}.
The acceptance in the cMSSM is similar, but different, from that in the
{\it Jets+mHT} selection.   The similarity is somewhat surprising, given that
$\alpha_T$ focuses on the two leading jets, and their relative angle, which is
characteristic of $\sq\sq^{*}$ production.\footnote{$\alpha_T$ handles the
case of multiple jets by combining them into two pseudo-jets, whose kinematics
are used in the naive two-jet formula.  This recombination helps recover some
jet energy lost to final state radiation.}  
The acceptance per process is shown in Fig.~\ref{fig:alfa2}.
There are patterns which are similar to those in Fig.~\ref{fig:classic2},
particularly the loss of efficiency near $\mgl\sim\msq$.

\begin{figure}[!ht]
\begin{center}
\includegraphics[width=0.75\textwidth]{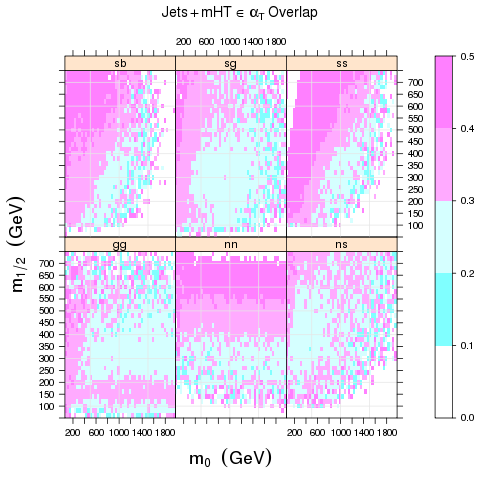}
\caption{
The fraction of events that pass both selections relative
to the events that pass the {\it Jets+mHT}.
}
\label{fig:both}
\end{center}
\end{figure}

Given the similarity of the exclusion limits (Fig.~\ref{fig:cmsSUSYCMSSM})
and the acceptance for the two selections, one wonders whether
the two analyses are highly correlated, despite the different cuts.
In Fig.~\ref{fig:both}, we address the overlap between the two
selections, by calculating the acceptance of events that pass
both selection requirements.   Here, we show the fraction of events
that pass both selections relative to those that pass the {\it Jets+mHT}
selection.
The overlap ranges from 20-50\% depending on the process and 
point in the $(m_0,\mhlf)$ plane.  There is less overlap
for $\gl\gl$ production and decay, and more for $\sq\sq$ topologies.
This suggests that a combination of these selections would be beneficial.

\begin{figure}[!ht]
\begin{center}
\includegraphics[width=0.75\textwidth]{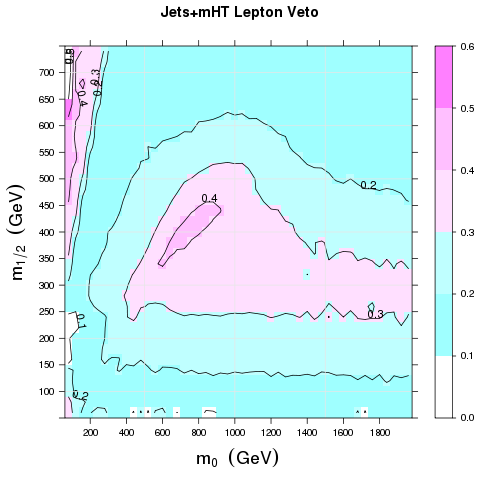}
\caption{
The fraction of events that fail the {\it Jets+mHT}
selection because of a veto on identified leptons.
}
\label{fig:lveto}
\end{center}
\end{figure}

There is one last issue regarding the hadronic analyses that we wish to address:
the impact of the lepton veto.  A veto is applied on events that have
an electron( muon) in the central rapidity region of the detector ($|\eta|<2.5$) 
with $p_T>10(15)$ GeV to reduce backgrounds.   In Figure~\ref{fig:lveto},
we show the fraction of events that fail the {\it Jets+mHT} selection
when the lepton veto is applied at the end.
The impact of the veto can be severe.  The increase near
$m_0=900$ GeV and $\mhlf=400$ GeV can be attributed to the
turn-on of the decay $\gl\to \widetilde{t}+\bar{t}$, where
the $\widetilde{t}$ can also decay to a $t$.
As we gain confidence in our modeling of Standard Model backgrounds,
we may be able to apply more sophisticated cuts
that reduce the
backgrounds without sacrificing signal.   This particular example
also illustrates
that simplified models that are too simple may not be able to catch all the
complexities of a particular model. 

\section{Other Phenomenological Considerations}

One may well wonder why $\sq\sq$ production and decay
passes an event selection requiring
3 or more jets, or why $\gl\gl$ production is not
constrained at the same level.
The presence of additional jet activity can arise from the
evolution of partonic structure through initial or final state radiation
(ISR or FSR).  The probability that a quark produced in the decay of 
a heavy squark at the scale $Q_F\sim \msq$ will remain a single parton
down to the jet resolution scale is given by the Sudakov form
factor, approximately:
$$\exp\big(-\frac{\alpha_s}{2\pi}C_F \ln^2\big(\frac{\msq^2}{p_T^2}\big)\big).$$
Choosing $\msq=500$ GeV and $p_T=50$ GeV, we estimate this probability at
roughly 40\%, or 16\% when considering a pair of squarks.
Clearly, vetoing a third jet would significantly reduce the signal acceptance.
The modeling of FSR, while not free from theoretical ambiguity, has been
studied in some detail at LEP (though for a color singlet system) and in
top decays at the Tevatron.  The estimate of ISR is less under control,
because of the dependence on parton distribution functions and the choice
of phase space for parton emission.   Jets from ISR will tend to be 
uncorrelated with the produced sparticles, whereas a selection
that includes jets from FSR might recover some of the resonant structure
in the signal process.

\begin{figure}
\begin{center}
\includegraphics[width=0.65\textwidth]{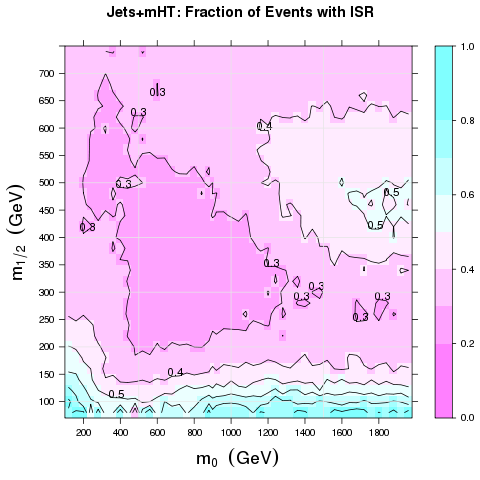}
\end{center}
\caption{The fraction of events that pass the {\it Jets+mHT} selection with
an additional jet from ISR in the cMSSM plane $(m_0,\mhlf)$ for $\tan\beta=10$.}
\label{fig:cmssmisr}
\end{figure}

Figure~\ref{fig:cmssmisr} shows the fraction of events at each $(m_0,\mhlf)$ point
that pass the {\it Jets+mHT} selection with an additional jet from ISR.   The origin
of a jet is traced using Monte Carlo truth information and the highest-$p_T$ 
quark or gluon constituent of a jet.
Typically 20-30\% of the jets arise from ISR.  The region where the gluino is
excluded relies more significantly on ISR, presumably because of the softening
of the jet spectra and $\met$ through the 3-body decay $\gl\to q\bar{q}\none$.

\begin{figure}
\begin{center}
\includegraphics[width=0.5\textwidth]{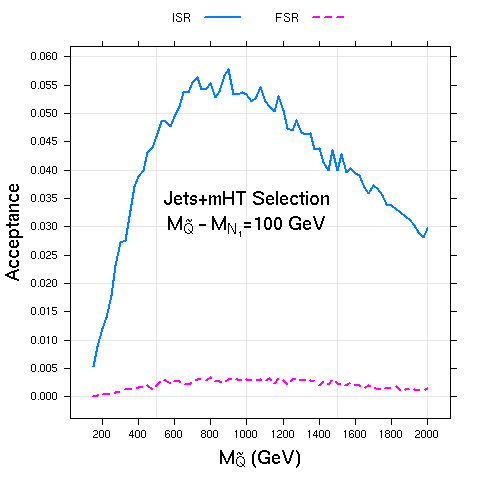}
\includegraphics[width=0.5\textwidth]{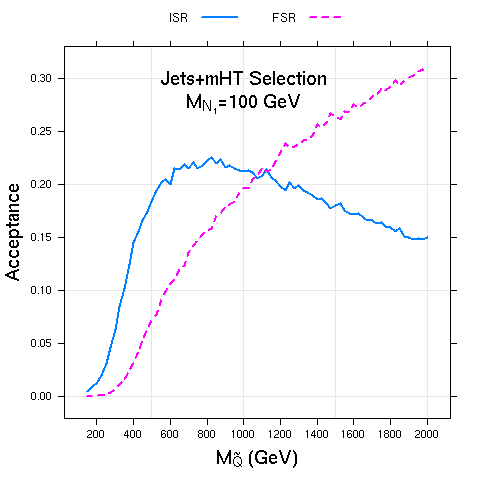}
\end{center}
\caption{The acceptance of events passing the {\it Jets+mHT} event selection
divided by their dependence on an ISR jet, for a fixed mass and varying
mass splitting.}
\label{fig:genisr}
\end{figure}

The cMSSM has typically large intrinsic $\met$, because of the large
mass splitting between the $\sq/\gl$ and $\none$.  
Also, the existence of cascade decays leads to additional hard jets in the final state.
This ameliorates some of
the dependence on ISR.
On the other hand,
more general spectra, which lead to different mass splitting, will be more
or less sensitive to ISR.
Figure~\ref{fig:genisr} shows the effect of the mass splitting for $\sq\sq$ events
that pass the {\it Jets+mHT} event selection.  Only the decay $\sq\to q\none$ is 
allowed.
The upper plot, with a fixed
mass splitting of 100 GeV, shows a strong dependence on ISR.   The lower plot,
where $\mno$ is fixed at 100 GeV, shows a more balanced dependence on ISR and FSR.
Some regions of Fig.~\ref{fig:cmssmisr} which overlap with the parameters in
the lower part of Fig.~\ref{fig:genisr} have $\sq$ cascade decays, 
such as $\sq\to q\ntwo(\to q\bar q\none$), and thus less dependence on in ISR.
Cascade decays, however, are more likely to produce at least one lepton in 
the final state, leading to a loss of acceptance with a leptonic veto.

The issue of the impact of ISR and FSR on high-$p_T$ searches
needs to be studied carefully.
To date, some basic studies have been performed \cite{Plehn:2005cq,Corke:2010zj,Alwall:2008va},
but these do not address theoretical uncertainties.
Currently, the systematic uncertainties
assessed on the experimental limits, by increasing or decreasing the amount of radiation,
are conservative.   A better description could allow us to set more stringent limits
on new particle production, or can be leveraged as a tool for discovery.
Data driven techniques would be beneficial,  perhaps based on observables
from $t\bar{t}$
production (particularly $p_T(t+\bar t)$).


\section{Conclusions \label{sec:conclusion}}

The textbook signature for {\sc SUSY} at hadron colliders
is {\it Jets+}$\met$, arising from the production and decay of the
gluino $\gl$ and the squarks $\sq$.      
An analysis of the limits arising from hadronic {\sc SUSY}
searches by CMS, and interpreted in the cMSSM, shows the current limits are
mainly excluding direct $\sq\sq$ and $\sq\gl$ production and decay.
The exclusion of $\sq\sq$ is an indirect exclusion of the the gluino
$\gl$ which mediates the process.  Particle-level
studies that follow the CMS event selections for the {\it Jets+mHT}
and $\alpha_T$ analyses, show that the dominant processes at the
cross selection level are also the dominant ones after cuts.
Our study also quantified the overlap between the two different
selections (20-50\%, depending on the process) and the impact
of the lepton veto (up to a 50\% loss of events).

The importance of $\sq\sq$ and $\sq\gl$ is interesting, since
interpretations of the data beyond the cMSSM often assume $\sq\sq^*$
and $\gl\gl$ production.  We suggest these new topologies and
ranges of parameters for them.   These processes need also to be
considered for a construction of a cMSSM point in terms of simplified models.

We also recognize the large potential of electroweak gaugino production
for exclusion or discovery of {\sc SUSY}.   Beyond the current limits,
chargino pair and neutralino-chargino production are the dominant processes.
These processes also impact the leptonic searches for {\sc SUSY}.
In fact, the current CMS leptonic limits are bordering the region of the
cMSSM plane (for the $\tan\beta=10$ slice) where the second heaviest neutralino decays
to a Higgs boson and the lightest neutralino. 

Because of the relevance of $\sq\sq$ production to the current limits,
we studied in some detail the importance of initial state radiation for
passing the {\sc SUSY} search selections.  For the cMSSM points considered,
final state radiation is more relevant, except for gluino masses less
than $300-400$ GeV.

\section{Appendix}

\subsection*{Squark-Gluino Reference Plot}

To ease the interpretation of figures displayed in
the $(m_0,\mhlf)$ plane, we collect here reference plots
displaying contours of physical masses.
\begin{figure}[!ht]
\begin{center}
\includegraphics[width=0.75\textwidth]{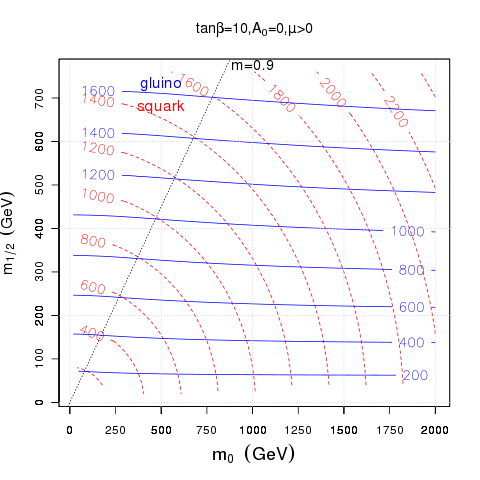}
\caption{
For reference, contours of squark and gluino masses in the cMSSM
for $\tan\beta=10, A_0=0~{\rm GeV}, sgn(\mu)>0$.   A line with slope
$m=0.9$ has been superimposed to show the approximation correlation
between $\msq$ and $\mgl$.
}
\label{fig:squark-gluino}
\end{center}
\end{figure}

\begin{figure}[!ht]
\begin{center}
\begin{tabular}{cc}
\includegraphics[width=0.5\textwidth]{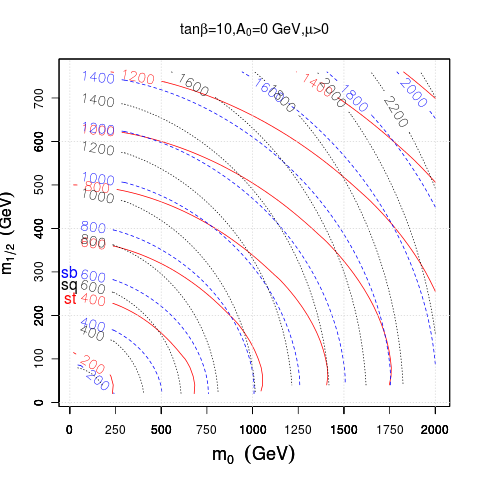}
\includegraphics[width=0.5\textwidth]{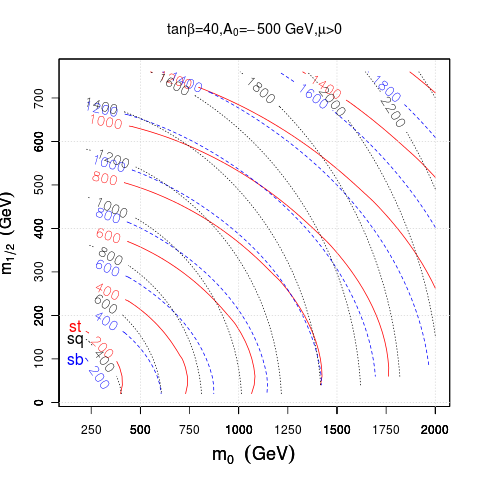}
\end{tabular}
\caption{
For reference, contours of sbottom, stop, and squark masses in the cMSSM
for $\tan\beta=10, A_0=0~{\rm GeV}, sgn(\mu)>0$ and
$\tan\beta=40, A_0=-500~{\rm GeV}, sgn(\mu)>0$.   
}
\label{fig:stop-sbottom-contour}
\end{center}
\end{figure}

\newpage

\subsection*{Dominant Processes for $\tan\beta=40$}
We present a similar plot as in Fig.~\ref{fig:frac2}, but
for a {\sc cMSSM} point with a large value of $\tan\beta$.
\begin{figure}[!ht]
\begin{center}
\begin{tabular}{cc}
\includegraphics[width=0.75\textwidth]{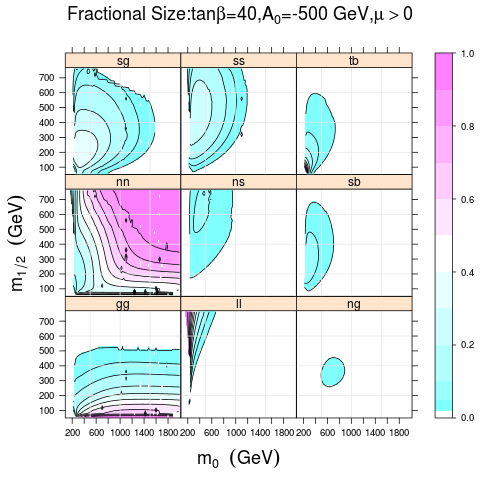}
\end{tabular}
\caption{
For reference, the fractional size of various subprocesses as a function
of ($m_0,\mhlf$) for $\tan\beta=40, A_0=-500~{\rm GeV}, sgn(\mu)>0$.
}
\label{fig:tanb40}
\end{center}
\end{figure}

\newpage

\subsection*{Squark-Squark and Squark-Antisquark Differences}
We provide additional information on the kinematic differences
between $\sq\sq$ and $\sq\sq^*$ production, namely the distribution
of the rapidity $y$ of the $\sq(\sq^*)$.  
\begin{figure}
\begin{center}
\includegraphics[width=0.65\textwidth]{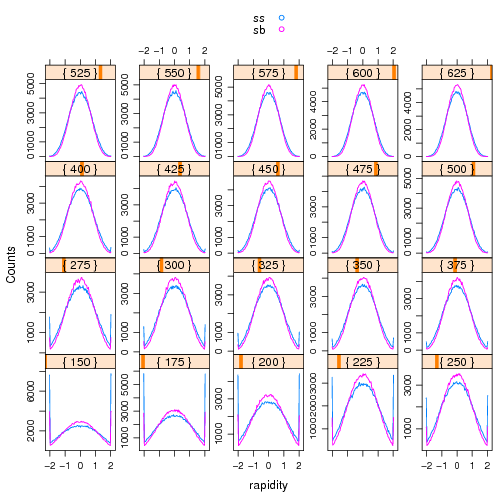}
\caption{  
Comparison of the rapidity $y$ 
of the produced $\sq,\sq^*$ for $\sq\sq$ versus $\sq\sq^*$ production.}
\end{center}
\end{figure}

\subsection*{Pseudo-Analysis Details}

Here, we provide details on the tools used to simulate the
detailed analyses performed by CMS, namely {\it Jets+mHT} and $\alpha_T$.
For each value of $(m_0,\mhlf)$ considered, the {\tt Pythia6}\cite{Sjostrand:2006za,Pythia6} 
Monte Carlo
was used to simulate the production, decay, and parton showering of 
sparticles.
The $p_T$-ordered shower was used, with the underlying-event and
hadronization effects turned off.
Information about the progeny of jets was based on 
Monte Carlo truth information about the highest $p_T$
parton inside the jet.  Jets were defined using 
the package {\tt FastJet}~\cite{Cacciari:2005hq,fastjet}.

The cuts applied are based on the description in the CMS analysis
studies.   The lepton veto is applied, again, using Monte Carlo truth
information and leptons from hard decays.

\section*{Acknowledgments}
The author thanks Seema Sharma for asking the correct questions
about the {\it Jets+mHT} analysis in CMS that inspired this
study, and to her, J. Lungu and J. Lykken for providing comments
on the draft.
S. Timm provided valuable help with running jobs on the Open Science
Grid.  
This research was done using resources provided by the Open Science Grid, 
which is supported by the National Science Foundation and the 
U.S. Department of Energy's Office of Science.
Fermilab is operated by the
Fermi Research Alliance, LLC under Contract
No. DE-AC02-07CH11359 with the United
States Department of Energy.


\end{document}